\documentclass[twocolumn]{aastex631}
\usepackage{graphicx,subfigure,color,amsmath,pifont,multirow,threeparttable}
\usepackage{acronym}
\usepackage{mathrsfs}
\usepackage{gensymb}
\usepackage{hyperref}

\shorttitle{Variabilities of Gamma-ray Bursts from the Dynamics of Fallback Material}
\shortauthors{Li et al.}
\begin{document}
\title{Variabilities of Gamma-ray Bursts from the Dynamics of Fallback Material after Tidal Disruption}
\correspondingauthor{Da-Bin Lin}
\email{lindabin@gxu.edu.cn}
\author{Yun-Peng Li}
\affiliation{Guangxi Key Laboratory for Relativistic Astrophysics, School of Physical Science and Technology, Guangxi University, \\Nanning 530004, China}
\author{Da-Bin Lin}
	\affiliation{Guangxi Key Laboratory for Relativistic Astrophysics, School of Physical Science and Technology, Guangxi University, \\Nanning 530004, China}
	\author{Guo-Yu Li}
	\affiliation{Guangxi Key Laboratory for Relativistic Astrophysics, School of Physical Science and Technology, Guangxi University, \\Nanning 530004, China}
	\author{Zi-Min Zhou}
	\affiliation{Guangxi Key Laboratory for Relativistic Astrophysics, School of Physical Science and Technology, Guangxi University, \\Nanning 530004, China}
	\author{En-Wei Liang}
	\affiliation{Guangxi Key Laboratory for Relativistic Astrophysics, School of Physical Science and Technology, Guangxi University, \\Nanning 530004, China}
	\begin{abstract}
		The gamma-ray burst (GRB) GRB~211211A and GRB~060614, believed to originate from the merger of compact objects,
        exhibit similarities to the jetted tidal disruption event (TDE) Sw~J1644+57,
        by showing violent variabilities in the light-curve during the decay phase.
        Previous studies suggest that such fluctuations in TDE may arise from the fallback of tidal disrupted debris.
        In this paper, we introduce the fluctuations of the mass distribution ${\rm d}M/{\rm d}E$ for the debris ejected during the tidal disruption (with energy $E$) and study their impact on jet power.
        Turbulence induced by tidal force and the self-gravity of the debris may imprint variabilities in ${\rm d}M/{\rm d}E$ during fallback.
        We model these fluctuations with a power density spectrum $\propto f_{\rm E}^{\beta}$, where $f_{\rm E} = 1/E$ and $\beta$ is the power-law index.
        We find that the resulting light curve can preserve the fluctuation characteristics from ${\rm d}M/{\rm d}E$.
        In addition, the observed fluctuations in the light-curves can be reproduced for a given suitable $\beta$.
        Based on the observations, we find that the value of $\beta$  should be around $-1$.
   \end{abstract}

	
\section{Introduction}\label{sec:intro}
Gamma-ray bursts (GRBs) are among the brightest transient phenomena in the Universe and have traditionally been classified into two categories based on the duration of their prompt emission. Long GRBs, lasting more than 2 seconds, are generally associated with the core collapse of massive stars and are often accompanied by supernovae; while short GRBs, with durations less than 2 seconds, are typically linked to the mergers of binary compact objects and accompanied by kilonovae (\citealp{1993ApJ...413L.101K,2006ARA&A..44..507W,2014ARA&A..52...43B,2015PhR...561....1K}).
However, growing observational evidence challenges this binary classification. For instance, GRB~060614 and GRB~211211A are long-duration GRBs that lack any associated supernovae (\citealp{2006Natur.444.1047F,2022Natur.612..232Y,2022Natur.612..228T}). Interestingly, these events exhibit characteristics commonly associated with short GRBs, such as evidence for accompanying kilonovae (\citealp{2015NatCo...6.7323Y,2022Natur.612..223R,2022ApJ...936L..10Z}). Kilonovae, powered by rapid neutron-capture (\textit{r}-process) nucleosynthesis following the merger of binary compact objects, strongly suggest that these GRBs likely originate from such mergers (\citealp{2006Natur.444.1044G,2006Natur.444.1053G,2022Natur.612..223R,2022Natur.612..228T,2022Natur.612..232Y,2022ApJ...933L..22Z}). However, the mechanism underlying their unexpectedly long durations remains poorly understood, posing challenges to the traditional classification of GRBs.

During the merger of neutron stars, tidal disruption of the neutron stars typically occurs (\citealp{2002MNRAS.334..481R,2007NJPh....9...17L,2007A&A...467..395O,2008PhRvD..77h4015S,2016MNRAS.460.3255R,2019MNRAS.485.4404D,2020ApJ...896L..38C}).
This implies that there are similarities in the physical mechanisms between short GRBs and tidal disruption events (TDE).
A TDE occurs when a star approaches a supermassive black hole (SMBH) and is partially or completely disrupted by tidal forces. Such events are predicted to result in the accretion of stellar debris by the SMBH, producing X-ray emission with typical decay timescales of months to years (\citealp{1975Natur.254..295H,1988Natur.333..523R,1989ApJ...346L..13E,2013ApJ...767...25G}). Notably, TDEs and short GRBs share key similarities: both involve tidal disruption process followed by debris fallback to form an accretion disk (\citealp{2013ApJ...767...25G,2015MNRAS.446..750F,2018ApJ...869..130R,2019GReGr..51...30S,2020PhRvD.101j3002K,2021ARA&A..59...21G,2024ApJ...968..104L}). The unusual $\gamma$-ray/X-ray transient Sw~J1644+57 is widely recognized as the first confirmed TDE driving a relativistic jet (\citealp{2011Natur.476..421B,2011Sci...333..203B,2011Sci...333..199L,2011Natur.476..425Z}). Intriguingly, the light curves of Sw~J1644+57, GRB~211211A, and GRB~060614 display similar features, including a $-5/3$ power-law decay following major initial pulses.
Moreover, the variabilities during the decay phase are also similar for these three transients,
suggesting a common underlying physical process.

The variability observed in the decay phases of these light curves offers critical insights into the dynamics of neutron star mergers, TDEs, and associated accretion processes.
The X-ray or $\gamma$-ray light curves of TDEs and the prompt emissions of gamma-ray bursts (GRBs) are widely believed to originate from internal dissipation processes within relativistic jets (\citealp{2007PhR...442..166N,2009ARA&A..47..567G,2018ApJ...859L..20D,2022ApJ...937L..12M}).
In TDEs, early initial pulses in the jet have been linked to jet precession or the early shocks near the pericenter, while the decay phases are attributed to fallback, as shown in previous studies (\citealp{2023ApJ...957L...9T,2024Natur.625..463S}).
We consider that the early main pulses of short GRBs originate from a physical process similar to that of jet TDEs.
A widely accepted framework for interpreting rapid temporal variability in GRB light curves is the internal shock model (\citealp{1994ApJ...430L..93R,1997ApJ...490...92K,1998MNRAS.296..275D,2009A&A...498..677B}). According to this model, the variability in the prompt emission reflects the central engine's activity history. This motivates an in-depth examination of how accretion process and fallback process dynamics contribute to light curve variability.
In this study, we analyze the variability characteristics in the decay phases of these transient events, uncover and investigate the physical mechanisms driving these variabilities.

Our paper is organized as follows. In Section~\ref{sec2}, we present fluctuation model within fallback process to simulate light curves. Section~\ref{sec3} show the results of the model and compare them with observations. Conclusions and discussions are provided in Section~\ref{sec4}.

\section{Method} \label{sec2}	
\subsection{Evolution of an Accretion Disk and the Corresponding Jet Power} \label{sub2:1}
    The debris ejected during the tidal disruption would fall back to the central engine and form an accretion disk around the compact object. Involving the contribution of fallback debris, the conservation of mass and angular momentum of an accretion disk can be described as follows (\citealp{2008bhad.book.....K}):
    \begin{equation} \label{eq1}
		\frac{\partial \Sigma}{\partial t} = \frac{1}{2\pi r} \frac{\partial \dot{M}}{\partial r} +\dot{\Sigma}_{\rm fb},
	\end{equation}
    \begin{equation}\label{eq2}
    	\frac{\partial}{\partial t}\left( r^{2}\Sigma\Omega \right) 
    	+ \frac{1}{r}\frac{\partial}{\partial r}\left( r^{3}\Sigma\Omega v_{\rm r} \right) 
    	= \frac{1}{r}\frac{\partial}{\partial r}\left( r^{3}\nu \Sigma\frac{\partial \Omega}{\partial r} \right) + r^{2}\Omega \dot{\Sigma}_{\rm fb},
    \end{equation}
   where $\Sigma(r)$ is the surface density of the disk at radius $r$ relative to the central black hole,
   $\dot{\Sigma}_{\rm fb}$ is the increase of the surface density due to the fallback of debris,
   $v_{\rm r}$ is radial inflow velocity of the disk and $\Omega$ is the angular velocity of the accretion disk material. Combining Equations~(\ref{eq1}) and~(\ref{eq2}), the accretion rate $\dot{M}$ can be derived as: 
	\begin{equation}\label{Eq:accretion_rate}
	\dot{M}(r,t)=6\pi\sqrt{r}\frac{\partial }{\partial r}(\nu\Sigma r^{1/2}).
	\end{equation}
	Then, the evolution of an accretion disk can be given by Equation~(\ref{eq1}) and~(\ref{Eq:accretion_rate}) \footnote{We verify the conservation of angular momentum by comparing the total angular momentum in the flow resulting from the star's disruption with that in the disk. We find that the total angular momentum in the flow is slightly larger than that in the disk, which suggests that our calculation process satisfies the conservation of angular momentum.}. 
In general, the kinematic viscosity $\nu$ is modelled as $\nu=\alpha c_{\rm{s}}H$, where the viscosity parameter $\alpha=0.1$ is adopted,
$c_{\rm{s}}$ is the sound velocity of gas, and $H$ is the half thickness of the disk.
	The outflow is not considered in Equation~$(1)$, as it is thought not to alter the highly variable behavior in AGN and the central engine of GRB (\citealp{2012MNRAS.421..308L,2016MNRAS.463..245L}).
	The values of $c_{\rm{s}}$ and $H$ are consistent with the advection factor $f_{\rm{adv}}(=Q_{\rm{adv}}^-/Q_{\rm{adv}}^+)$ of the accretion flow, i.e. $c_{\rm{s}}\sim v_{\rm{\phi}}\sqrt{f_{\rm{adv}}}$ and $H/r\approx \sqrt{f_{\rm{adv}}}$, where $Q_{\rm{adv}}^-$ and $Q_{\rm{adv}}^+$ are the factors of the advection cooling and viscous heating in the accretion flow, and $v_{\rm{\phi}}$ is the Kepler’s rotation velocity of gas around black hole.
	Since the value of $H$ does not change significantly for the advection-dominated accretion flow (ADAF, $f_{\rm{adv}}\sim 1$; \citealp{1994ApJ...428L..13N}) and neutrino-cooling-dominated accretion flow (NDAF, $f_{\rm{adv}}\gtrsim 0.01$; \citealp{1999ApJ...518..356P,2005ApJ...629..341K}), $H/r=0.5$ and $c_{\rm{s}}= v_{\rm{\phi}}/2$ are adopted in our calculation.

    The formation of a jet is associated with the accretion.
    The dominant paradigms for jet production are outlined in the works of \cite{1977MNRAS.179..433B} and \cite{1982MNRAS.199..883B}.
	In this paper, the jet power is simply set to be proportional to the black hole's accretion rate as
	\begin{equation}
	L_{\rm jet}=\eta_{\rm jet}\eta_{\rm acc} \dot{M}_{\rm{in}}c^2,
	\label{f5}
	\end{equation}
	where $\eta_{\rm acc}$ is the energy conversion efficiency of the accreting material and $\eta_{\rm acc}=0.1$ is taken (\citealp{2004MNRAS.351..169M,2004MNRAS.354.1020S}), $\eta_{\rm jet}$ is the proportion of the total released accretion energy used to power the jet,
    $c$ is speed of light, and $\dot{M}_{\rm{in}}$ is the accretion rate of the innermost annulus of the disk.

For reasonable values of $\alpha$, the viscous timescale at the tidal radius is significantly shorter than the characteristic fallback time at later time. As a result, the accretion rate effectively follows the fallback rate (\citealp{1988Natur.333..523R,1990ApJ...351...38C}). By explicitly parameterizing $\alpha$, we find that fluctuations in the accretion flow induced by $\alpha$ variations do not lead to significant variability in the jet luminosity. Consequently, we focus on the variability of the jet luminosity driven by fluctuations in the fallback process.

\subsection{Fluctuations in the Fallback Process} \label{sub3:1}
	The tidal disruption of a compact binary or massive star would eject a large amount of debris, part of which remains bound and falls back to form an accretion disk.
    These bound debris moves along Keplerian orbits, with mass distribution ${\rm d}M/{\rm d}E$ in energy space.
	The fallback mass rate as a function of time $t$ can then be expressed as (\citealp{2009MNRAS.392..332L,2013ApJ...767...25G})
	\begin{equation}\label{Eq:M_fb}
		\dot{M}_{\rm fb}=\frac{{\rm d}M}{{\rm d}t} = \frac{{\rm d}M}{{\rm d}E}  \left| \frac{{\rm d}E}{{\rm d}t} \right|=\frac{(2\pi GM_{\rm BH})^{2/3}}{3}\frac{{\rm d}M}{{\rm d}E}t^{-5/3},
	\end{equation}
    where $E=-({2\pi GM_{\rm BH}}/{t})^{2/3}/2$ is the specific orbital energy of the bound debris, $M_{\rm BH}$ is the mass of the black hole, and $G$ is the gravitational constant. Equation~(\ref{Eq:M_fb}) reveals that the fallback rate $\dot{M}_{\rm fb}$ and its fluctuations are directly associated with the mass distribution in the energy space ${\rm d}M/{\rm d}E$.	
	In this study, $\overline{{\rm d}M/{\rm d}E}$ is defined as the average value of the mass distribution of fallback material in energy space, neglecting the effects of fluctuations. 
    Hydrodynamic studies (\citealp{2009MNRAS.392..332L,2013ApJ...767...25G,2023ApJ...946...25J}) and numerical calculations (\citealp{1989ApJ...346L..13E}) suggest that $\overline{{\rm d}M/{\rm d}E}$ is a function of $|E|$, remaining approximately constant for $|E|<|E_{\rm c}|$ and declining steeply for $|E|>|E_{\rm c}|$. For $|E|<|E_{\rm c}|$, the fallback material returns to the accretion disk following a steady decay ($\dot{M}_{\rm fb} \propto t^{-5/3}$) after complete tidal disruption. Therefore, $E_{\rm c}$ corresponds to the peak time of the light curve, $t_{\rm peak}$ (\citealp{2019GReGr..51...30S}).
    Then, we use the following analytical form to describe $\overline{{\rm d}M/{\rm d}E}$ , i.e.,
	\begin{equation}
		\overline{\frac{{\rm d}M}{{\rm d}E}} =
		\begin{cases}
			\displaystyle \frac{M_{\rm fb}}{|E_{\rm c}|-|E|_{\rm min}}, &  |E|\le |E_{\rm c}|, \\[10pt]
			\displaystyle \frac{M_{\rm fb}}{|E_{\rm c}|-|E|_{\rm min}} \exp\left[-\frac{(|E|-|E_{\rm c}|)^2}{2\sigma^2}\right], &  |E| > |E_{\rm c}|.
			
		\end{cases} \label{eq5}
	\end{equation} 
Here, $|E|_{\rm min}$ denotes the minimum energy of the fallback material, $M_{\rm fb}$ is the total mass of fallback material within the energy range $|E|_{\rm min}< |E|\le |E_{\rm c}|$, and $\sigma$ characterizes the decline of $\overline{{{\rm d}M}/{\rm d}E}$ at $|E| > |E_{\rm c}|$. In this work, we adopt $\sigma = \left| E_{\rm c} \right| / 2$ and $M_{\rm fb} = 0.1 M_{\odot}$. Equation~(\ref{eq5}) thus reproduces both the rising and decaying phases of the fallback light curve, consistent with observed TDE light curves. (\citealp{2021ARA&A..59...21G}).

Previous studies have indicated that the distribution ${\rm d}M/{\rm d}E$ exhibits significant fluctuations due to various effects, such as self-gravity and shock heating (\citealp{2015ApJ...808L..11C,2020ApJ...896L..38C,2021ApJ...923..184N,2023MNRAS.526.2323F}). 
In addition, we think that other factors, such as the star's non-uniform density distribution and turbulence during tidal disruption, also play important roles in generating these fluctuations.
In this paper, the fluctuation of ${\rm d}M/{\rm d}E$ in energy $E$ space is modelled as
\begin{equation} \label{eq77}
\frac{{\rm d}M}{{\rm d}E}=\overline{\frac{{\rm d}M}{{\rm d}E}}[1+b_{\rm E}u_{\rm E}(E_{\rm cut}, E)]^{\xi},
\end{equation}
where $b_{\rm E} (< 1)$ and $\xi$ are constants.
Considering the properties of turbulence (\citealp{2002PhFl...14.1065G}),
the variability of $u_{\rm E}(E_{\rm cut}, E)$ is modeled with a power density spectral (PDS) as
\begin{equation}\label{eq7}
P_{\rm Ef} \propto \frac{2Qf_{\rm cut}f_{\rm E}^{\beta}}{f_{\rm cut}^2+4Q^2(f_{\rm E}-f_{\rm cut})^2},
\end{equation}
where $\beta\le0$ is a constant and $f_{\rm cut} = 1/E_{\rm cut}$ is the cutoff frequency for the fluctuations in the fallback.
The cutoff energy $E_{\rm cut}$ is the characteristic energy
of which the fluctuations is suppressed.
$\left| E\right |_{\rm min}$ represents the minimal binding energy among the fallback materials and thereby setting a natural threshold below which fluctuations are reduced.
In this paper, we take $E_{\rm cut}=\eta_{\rm E} \left| E\right |_{\rm min}$, with $\eta_{\rm E}$ being a constant.
Here, $P_{\rm Ef}$ is the Fourier transform of $u_{\rm E}(E_{\rm cut}, E)$ and $Q=0.5$ is the quality factor, which is equal to the ratio of the Lorentz peak frequency to the full width at half maximum (\citealp{2006MNRAS.367..801A,2016MNRAS.463..245L}).
It should be noted that the frequency in energy space is defined as $f_{\rm E} \propto 1/E$, analogous to the time-domain frequency $f \propto 1/t$. As a result, low-energy components contribute less power in Equation~(\ref{eq7}), leading to a suppression of light curve fluctuations at later times. Consequently, fallback debris with lower energy is less affected by perturbations.

The material is assumed to fall onto the disk at around the radius $r_{\rm fb}$.
The spatial distribution of the fallback material is assumed as
\begin{equation}
\dot{\Sigma}_{\rm fb}=\frac{\dot{M}_{\rm fb}}{2\pi r}A\exp\left[-(\frac{r-r_{\rm fb}}{r_{\rm fb}/4})^2\right],
\end{equation}
where the normalization factor $A$ is obtained by setting $\int_{r_{\rm in}}^{r_{\rm out}}{\dot\Sigma 2\pi r dr}  = {\dot M_{{\rm{fb}}}}$.

\section{Result} \label{sec3}

\subsection{Variabilities of the Jet Power} \label{sub3:2}

\begin{figure*}[htb]
	\centering
	\begin{minipage}{0.49\linewidth}
		\includegraphics[height=0.7\textwidth, width=1\textwidth]{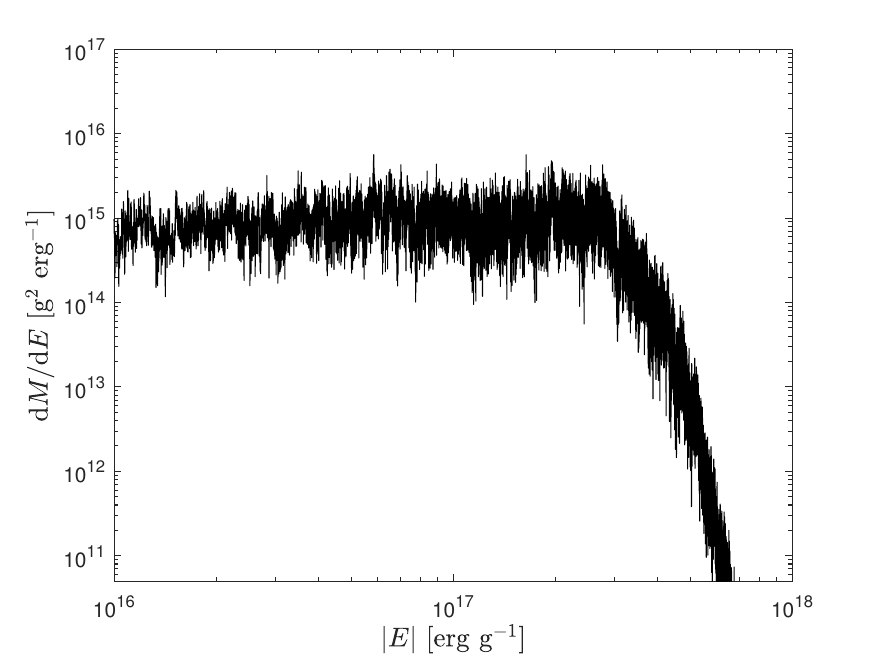}
	\end{minipage}
	\begin{minipage}{0.49\linewidth}
		\includegraphics[height=0.7\textwidth, width=1\textwidth]{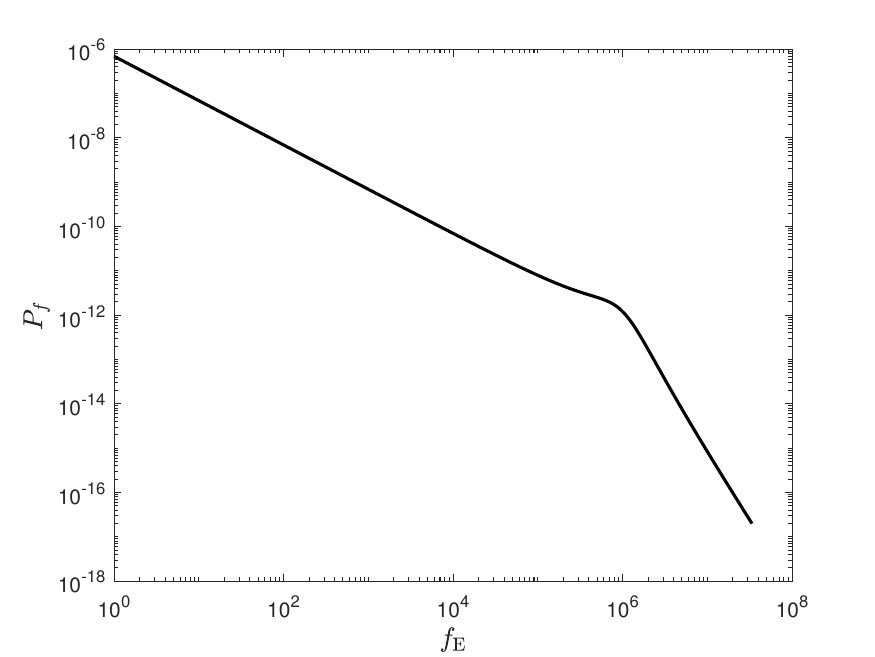}
	\end{minipage}
	
	\caption{
	The distribution of ${\rm d}M/{\rm d}E$ in energy space with fluctuations introduced (left) and the shape of the corresponding power density spectrum (PDS) (right), with parameters given by $(Q,\eta_{\rm E},\beta,\xi,b_{\rm E},t_{\rm peak}) = (0.5,0.01,-1,5,0.7,3\,\mathrm{s})$.
	}

	\label{fig1}
\end{figure*}

\begin{figure*}[htb]
\centering

			\includegraphics[height=0.7\textwidth, width=1\textwidth]{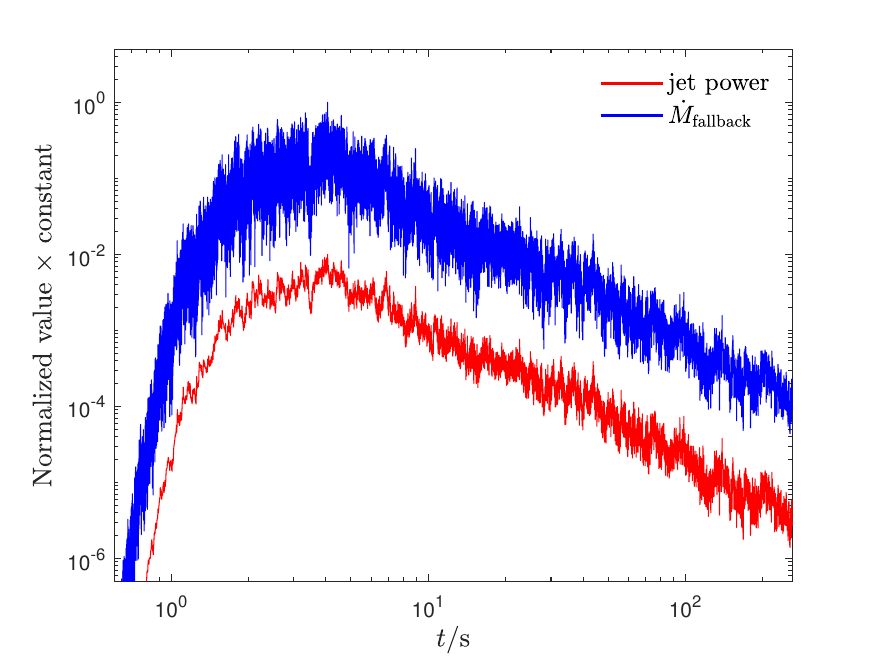}

	\caption{Comparison of the fallback rate (blue) and jet power (red) predicted by ${\rm{d}} M/{\rm{d}} E$ fluctuation models. To prevent overlap, the normalized curves are scaled by factors of 1 (blue) and 0.01 (red). Both curves correspond to the case of $(Q,\eta_{\rm E},\beta,\xi,b_{\rm E},t_{\rm peak})=(0.5,0.01,-1,5,0.7,3\,\mathrm{s})$.}

		\label{fig2}
	\end{figure*}

We incorporate fluctuations in the fallback process to examine the variability characteristics of the light curve.
According to Equations~(\ref{eq77}) and~(\ref{eq7}), we first present the $\mathrm{d}M/\mathrm{d}E$ fluctuations in energy space and the shape of the corresponding PDS, with parameters $(Q,\eta_{\rm E},\beta,\xi,b_{\rm E},t_{\rm peak}) = (0.5,0.01,-1,5,0.7,3\,\mathrm{s})$, as shown in Figure~\ref{fig1}. In the left panel of Figure~\ref{fig1}, $\mathrm{d}M/\mathrm{d}E$ fluctuates strongly around the average distribution $\overline{\mathrm{d}M/\mathrm{d}E}$ given by Equation~(\ref{eq5}). The amplitude of these fluctuations increase with energy. 
Given the $\mathrm{d}M/\mathrm{d}E$ distribution, the fallback rate $\dot{M}_{\rm fb}$ is calculated from Equation~(\ref{Eq:M_fb}).
In what follows, we set the fallback radius $r_{\rm fb}$ equal to the tidal disruption radius $r_{\rm t}$. The tidal disruption radius of a neutron star can be estimated as $r_{\rm t} = R_{\rm NS}(M_{\rm BH}/M_{\rm NS})^{1/3}$ (\citealp{2019GReGr..51...30S,2021LRR....24....5K}), where $R_{\rm NS}$ is the neutron star radius, $M_{\rm NS}$ is its mass, and $M_{\rm BH}$ is the mass of the central black hole. Here, we adopt $R_{\rm NS}=12\,{\rm km}$, $M_{\rm NS}=1.4\,M_{\odot}$, and $M_{\rm BH}=3\,M_{\odot}$.
In Figure~\ref{fig2}, we illustrate the corresponding temporal evolution of the fallback rate (blue curve) and the jet power (red curve) for a case with $(Q,\eta_{\rm E},\beta,\xi,b_{\rm E},t_{\rm peak})=(0.5,0.01,-1,5,0.7,3\,\mathrm{s})$.
One can find that the jet power preserves the general variability morphology of the fallback process
with short timescale fluctuations slightly suppressed.
This is owning to that the accretion process suppresses the short timescale fluctuations,
of which the timescale is shorter than the viscous timescale for the disk at the fallback radius $r_{\rm t}$.

\begin{figure*}[htb]
	\centering
    \begin{minipage}{0.49\linewidth}
		\includegraphics[height=0.7\textwidth, width=1\textwidth]{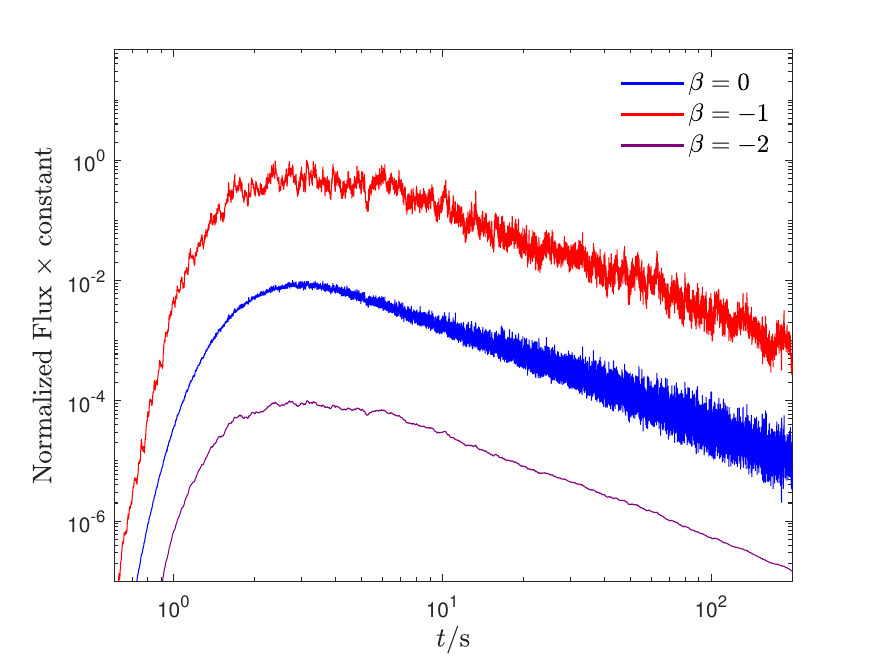}
	\end{minipage}
	\begin{minipage}{0.49\linewidth}
      \includegraphics[height=0.7\textwidth, width=1\textwidth]{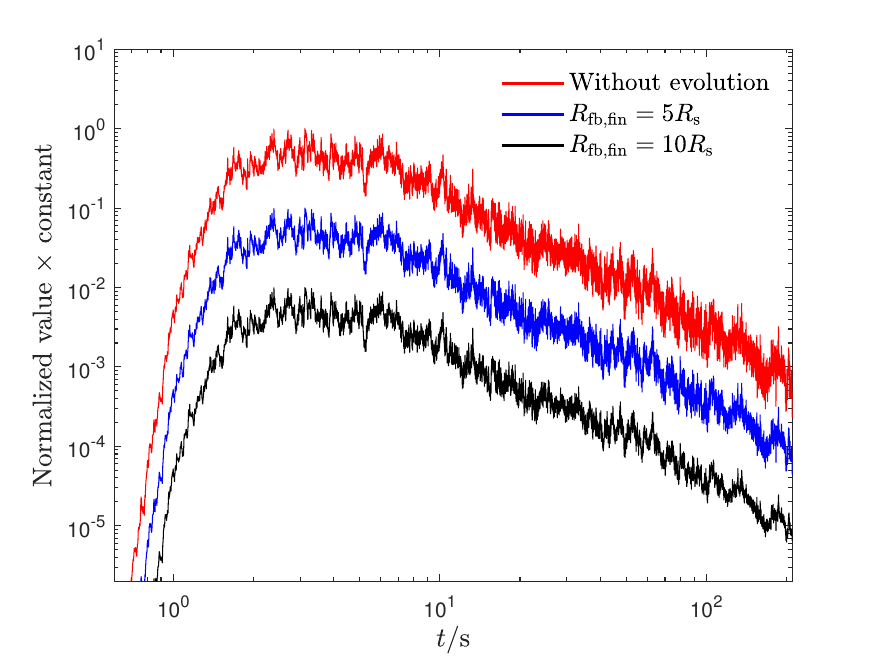}
	\end{minipage}
	
\caption{Normalized light curves for different fluctuation parameters ($\beta$) and final fallback radii ($r_{\rm fb,fin}$). 
The left panel displays normalized light curves for different fluctuation parameters $\beta$. To prevent overlap, the curves are scaled by factors of 1 (red), $10^{-2}$ (blue), and $10^{-4}$ (purple). These curves correspond to the case of $(Q,\eta_{\rm E},\xi,b_{\rm E},t_{\rm peak})=(0.5,0.01,5,0.7,3\,\mathrm{s})$, with $\beta = 0$ (blue), $\beta = -1$ (red), and $\beta = -2$ (purple).
The right panel shows normalized light curves for different final fallback radii. Similarly, the curves are scaled by factors of 1 (red), $10^{-2}$ (blue), and $10^{-4}$ (black) for clarity. All curves are generated with $(Q, \eta_{\rm E}, \beta, \xi, b_{\rm E}, t_{\rm peak}) = (0.5, 0.01, -1, 5, 0.7, 3\,\mathrm{s})$, corresponding to a static fallback radius (red), $r_{\rm fb,fin} = 5r_{\rm s}$ (blue), and $r_{\rm fb,fin} = 10r_{\rm s}$ (black), where $r_{\rm s} = 2GM_{\rm BH}/c^2$ is the Schwarzschild radius of the black hole.
}

    \label{fig3}
\end{figure*}

Next, we study the effects of different parameters on variabilities of the light curves.
In the left panel of Figure~\ref{fig3}, we plot the jet power for cases with different $\beta$, using the model parameters $(Q,\eta_{\rm E},\beta,\xi,b_{\rm E},t_{\rm peak}) = (0.5,0.01,-1,5,0.7,3\,\mathrm{s})$. We find that $\beta$ plays a crucial role in shaping the jet power fluctuations.
In particular, the case $\beta \approx -1$ closely reproduces the observed variability pattern of the jet power.
Since the disrupted debris initially spans a range of radii around the black hole, the fallback radius $r_{\rm fb}$ naturally evolves over time. In our simulation, we increase the fallback radius linearly with time from the tidal radius $r_{\rm t}$ to a final value $r_{\rm fb,fin}$ over the same duration. Figure~\ref{fig3} (right panel) illustrates that this temporal evolution of the fallback radius does not change the overall morphology of the light curve fluctuations, but slightly suppresses its amplitude as the fallback radius increases.

\begin{figure*}[htb]
	\centering
	
	\includegraphics[height=0.7\textwidth, width=1\textwidth]{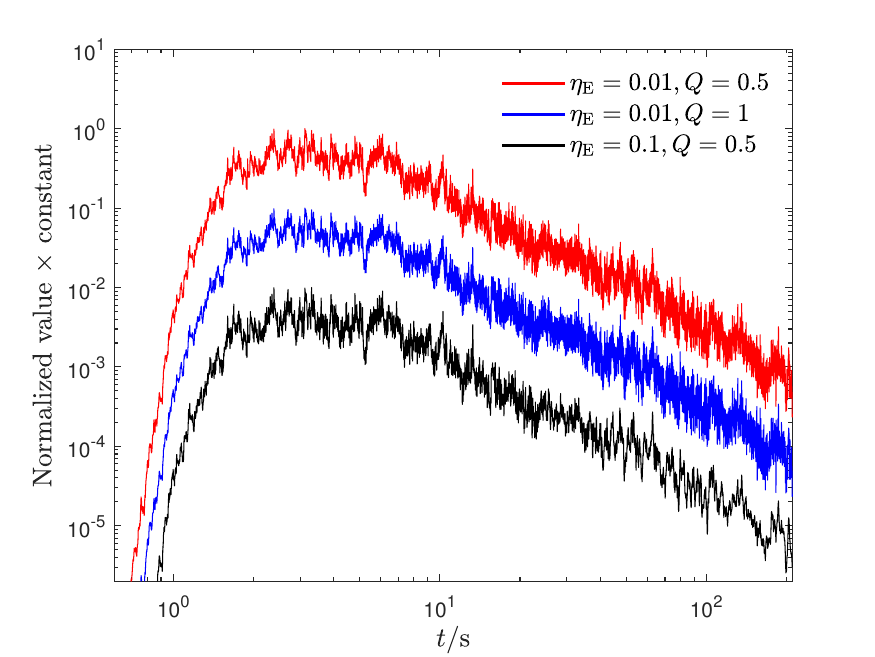}
	
	\caption{Normalized light curves for different parameters $\eta_{\rm E}$ and $Q$. To prevent overlap, the normalized curves are scaled by factors of 1 (blue), 0.1 (red) and 0.01 (blue). all of curves correspond to the case of $(\beta,\xi,b_{\rm E},t_{\rm peak})=(-1,5,0.7,3\,\mathrm{s})$.}

	\label{figend}
\end{figure*}

\begin{figure*}[htb]
	\centering
	\begin{minipage}{0.49\linewidth}
		\includegraphics[height=0.7\textwidth, width=1\textwidth]{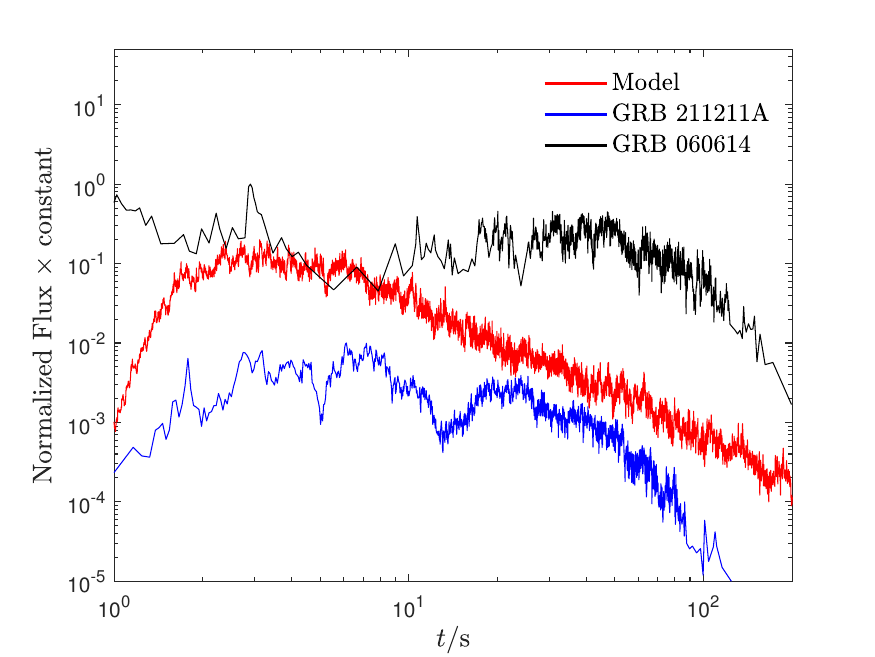}
	\end{minipage}
	\begin{minipage}{0.49\linewidth}
		\includegraphics[height=0.7\textwidth, width=1\textwidth]{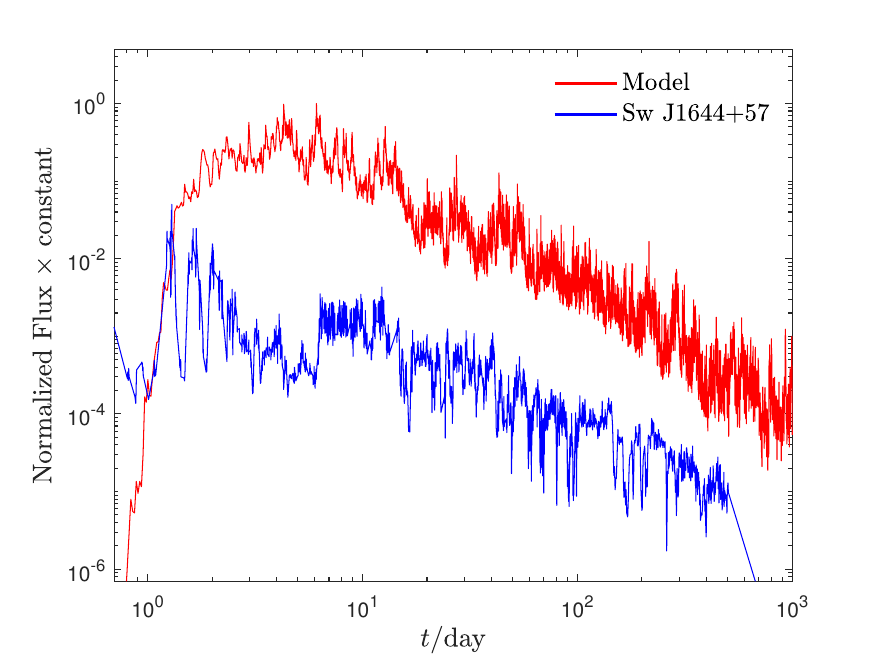}
\end{minipage}	
\caption{
Comparison of observed normalized light curves and those generated by ${\rm d}M/{\rm d}E$ fluctuation models for sGRBs (left panel) and TDEs (right panel). In the left panel, the curves are scaled by factors of 1 (black), 0.1 (red), and 0.01 (blue). The red solid line represents the model with parameters $(Q,\eta_{\rm E},\beta,\xi,b_{\rm E},t_{\rm peak})=(0.5,0.01,-1,5,0.5,3\,\mathrm{s})$, while the black and blue solid lines correspond to the observed light curves of GRB~060614 and GRB~211211A, respectively.
In the right panel, the curves are scaled by factors of 1 (red) and 0.1 (blue). The red solid line represents the model with parameters $(Q,\eta_{\rm E},\beta,\xi,b_{\rm E},t_{\rm peak}, M_{\rm BH}, M_{\rm star}, R_{\rm star})=(0.5,0.01,-1,10,0.8,5\,\mathrm{day}, 10^6M_{\odot}, 1M_{\odot}, 1R_{\odot})$, while the blue solid line corresponds to the observed light curve of Sw~J1644+57.}

    \label{fig34}
\end{figure*}

We further examine whether the fluctuation morphology of the light curve produced by the $dM/dE$ fluctuation model is affected by other parameters. We find that variations in the total fallback mass $M_{\rm fb}$ and the viscosity parameter $\alpha$ have negligible impact on the fluctuation characteristics of the light curve. The parameters $E_{\rm c}$ (which determines the peak time $t_{\rm peak}$) and $\sigma$ in Equation~(\ref{eq5}) affect only the peak position and the slope of the rising segment of the light curve, respectively, without altering the fluctuation morphology. Parameters such as $b_{\rm E}$ and $\xi$ influence only the amplitude of the fluctuations. The light curves plotted in Figure~\ref{figend} for different values of the $E_{\rm cut}$ and $Q$ show that a larger $E_{\rm cut}$ suppresses late-time fluctuations, but neither parameter changes the fluctuation morphology. Thus, the fluctuation characteristics of the light curve depend solely on the index $\beta$.  
As a demonstration, we plot the jet power variability for two parameter sets: $(Q,\eta_{\rm E},\beta,\xi,b_{\rm E},t_{\rm peak}) = (0.5,0.01,-1,5,0.5,3\,\mathrm{s})$ and $(Q,\eta_{\rm E},\beta,\xi,b_{\rm E},t_{\rm peak}, M_{\rm BH}, M_{\rm star}, R_{\rm star}) = (0.5,0.01,-1,10,0.8,5\,\mathrm{day},10^6\,M_{\odot},1\,M_{\odot},1\,R_{\odot})$. These cases are compared with the observed jet power variability of GRB~211211A, GRB~060614, and Sw~J1644+57 in Figure~\ref{fig34}. Here $M_{\rm star}$ and $R_{\rm star}$ denote the mass and radius of the disrupted star, respectively.

\begin{table}
	\centering{
		\caption{Fitting parameters of the PDSs}\label{tb1}
		\begin{tabular}{ccccccc}
			\hline \hline
			Event & ${\rm{log}}\,N$  &  $\alpha$   & ${\rm{log}}\,B$
			\tabularnewline
			\hline
			GRB~211211A & $-2.59 \pm 0.09$ & $-0.94 \pm 0.18$ & $-3.95 \pm 1.33$ \\
			GRB~060614  & $-2.76 \pm 0.17$ & $-1.30 \pm 0.28$ & $-2.93 \pm 0.13$ \\
			Sw~J1644+57 & $0.24 \pm 0.56$ & $-0.61 \pm 0.14$ & $2.72 \pm 0.03$ \\
			
			\hline
	\end{tabular}}
	
\end{table}

\subsection{Calculation and Fitting of Power Density Spectrum} \label{sub3:3}
\begin{figure*}[htb]
	\centering
	\includegraphics[width=0.32\textwidth]{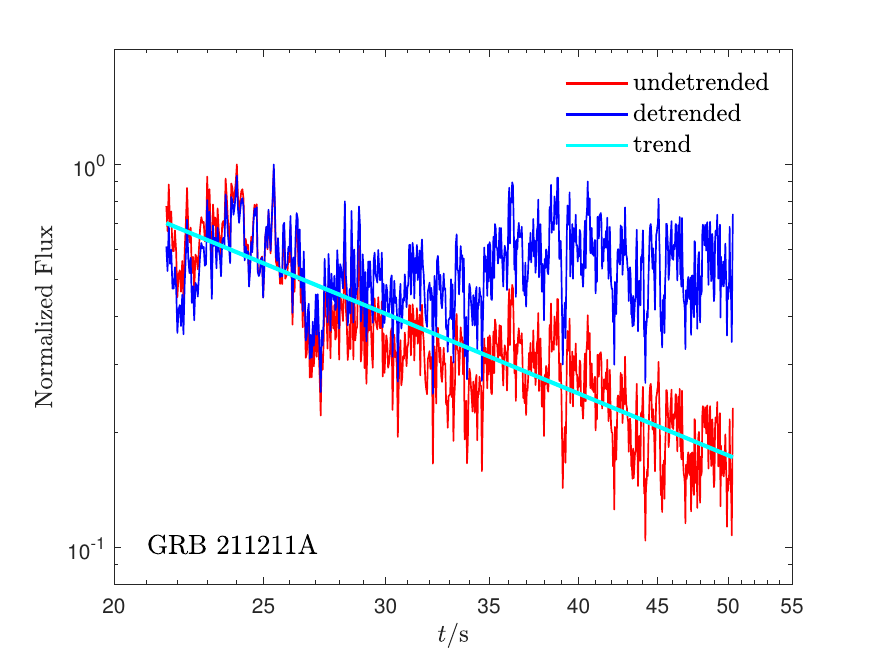}
	\includegraphics[width=0.32\textwidth]{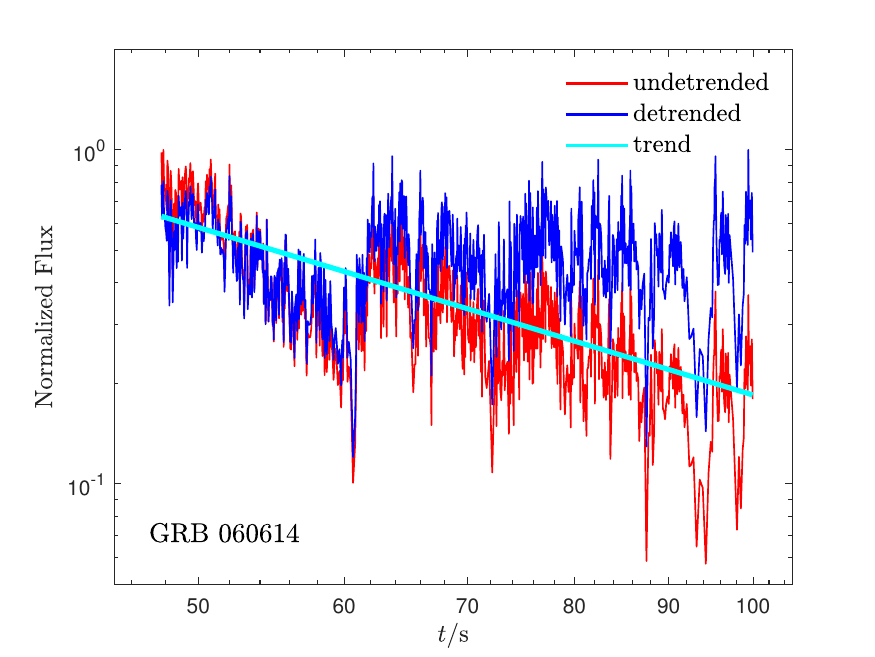}
	\includegraphics[width=0.32\textwidth]{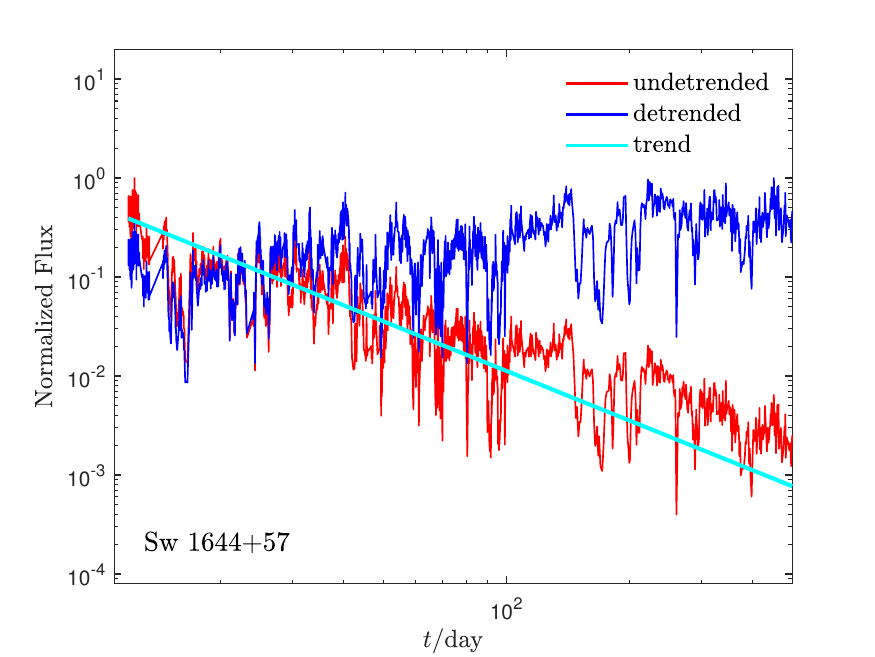}
	
	\caption{Comparison of normalized light curves before and after detrending. The left, middle, and right panels show light curves for GRB~211211A, GRB~060614, and Sw~J1644+57, respectively. The red solid lines represent the original light curves, the blue solid lines represent the detrended light curves, and the cyan solid lines represent the fitted trends of the light curves.}
	
	\label{fig44}
\end{figure*}

Figure~\ref{fig34} reveals that the fluctuations in the fallback process
can reproduce the variabilities in the observations.
In addition, the variabilities of the jet power effectively reserve the variabilities in the fallback process.
Then, we proceed by investigating the PDS of the observed light curves to examine the fluctuations in the fallback process.
We employed the Lomb–Scargle periodogram (LSP; for more details, see \citealp{1976Ap&SS..39..447L,1982ApJ...263..835S,2009A&A...496..577Z}) to calculate the PDS for GRB~211211A, GRB~060614, and Sw~J1644+57.
Here, the power-law decay segment in the light curve,
which is believed to be present in a tidal disruption event,
is selected for our fluctuating PDS analysis. 
To minimize the effect of the overall trend in the light curve on the PDS analysis,
we de-trend the overall trend for our selected segments (for further details, see Appendix).
Both the original and detrended light curves, along with the trend, are presented in Figure~\ref{fig44}. 

\begin{figure*}[htb]
	\centering
	\includegraphics[width=0.32\textwidth]{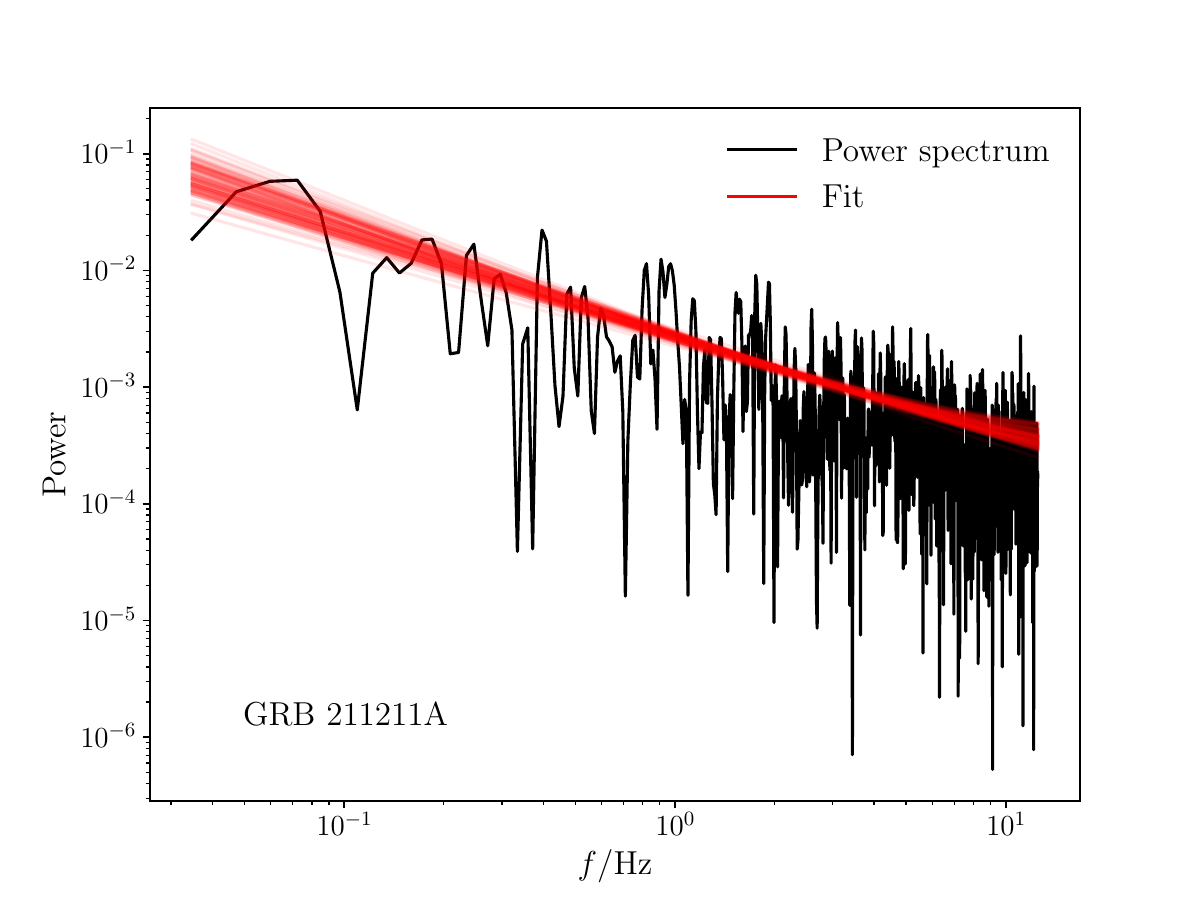}
	\includegraphics[width=0.32\textwidth]{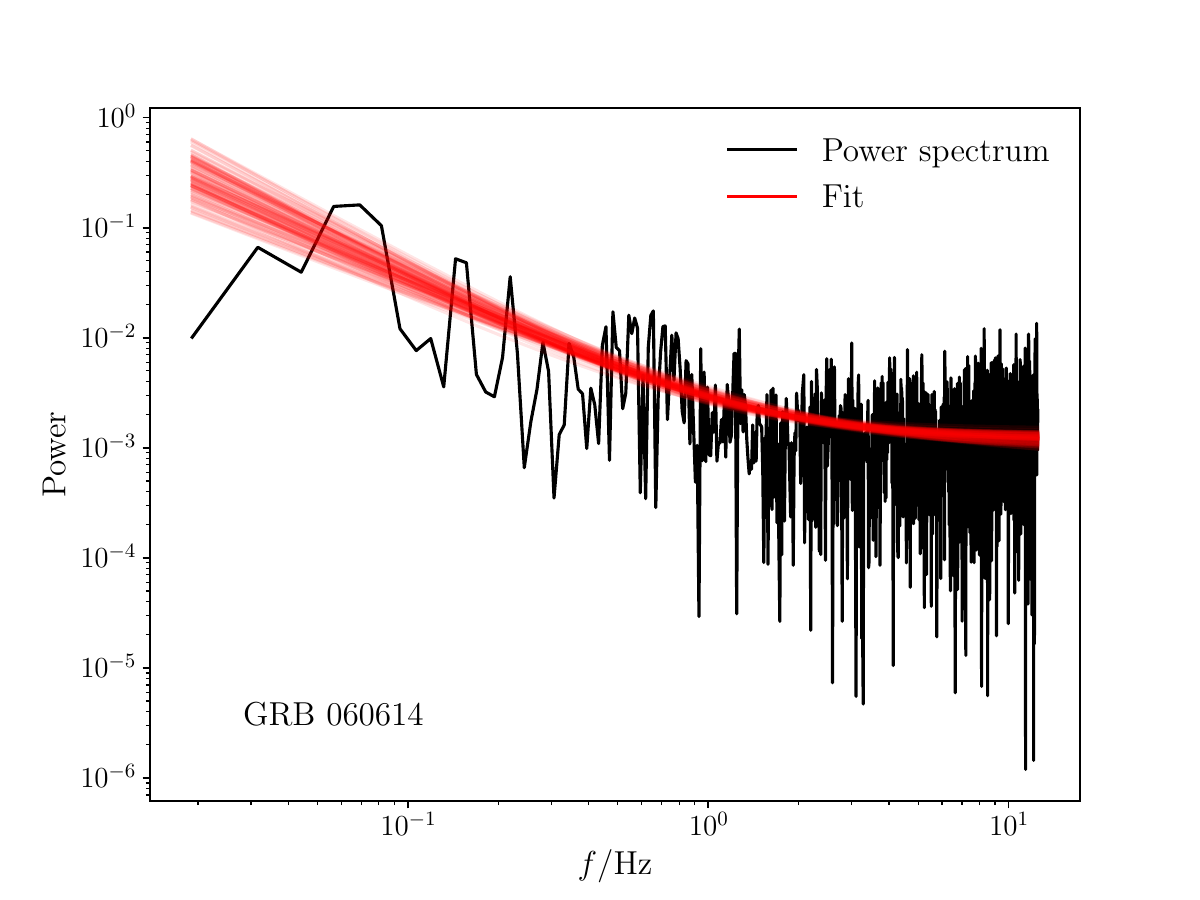}
	\includegraphics[width=0.32\textwidth]{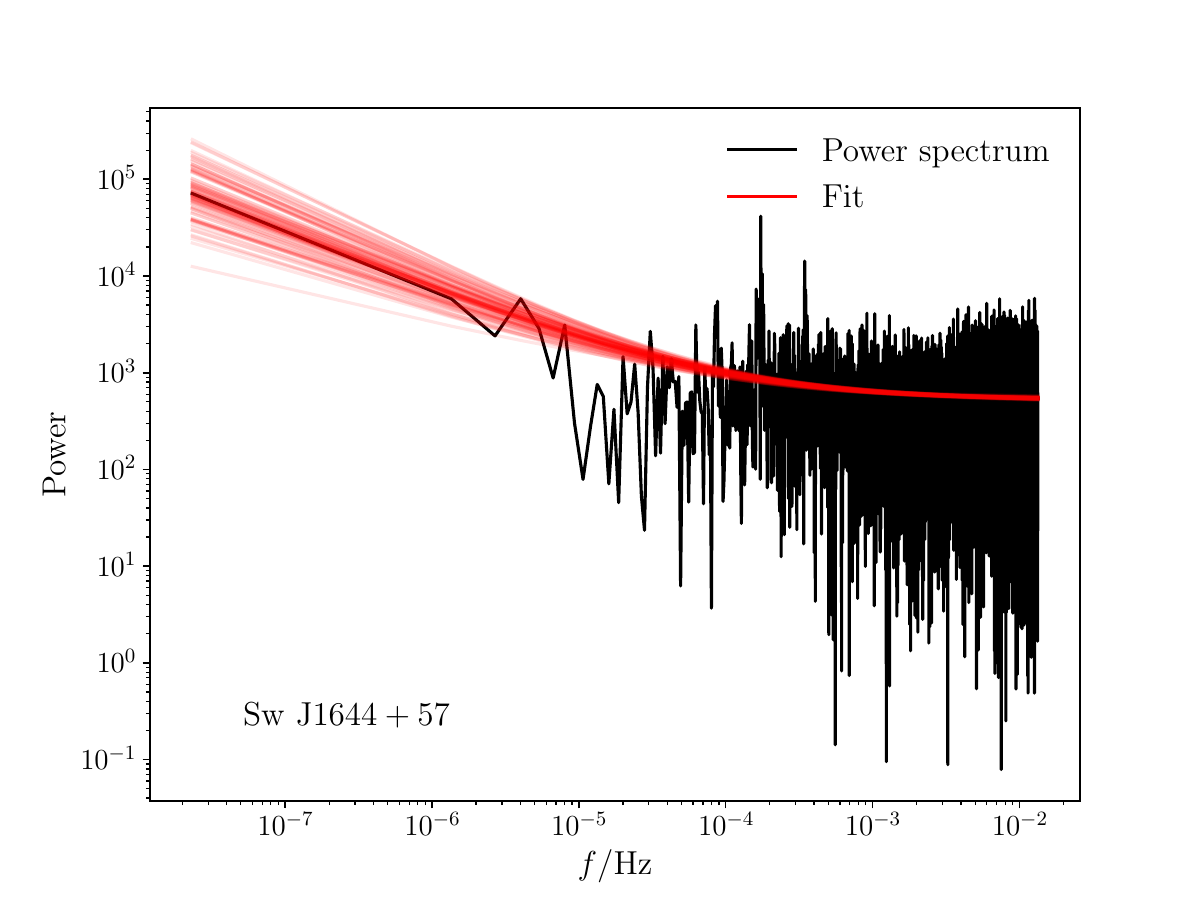}
	\includegraphics[width=0.32\textwidth]{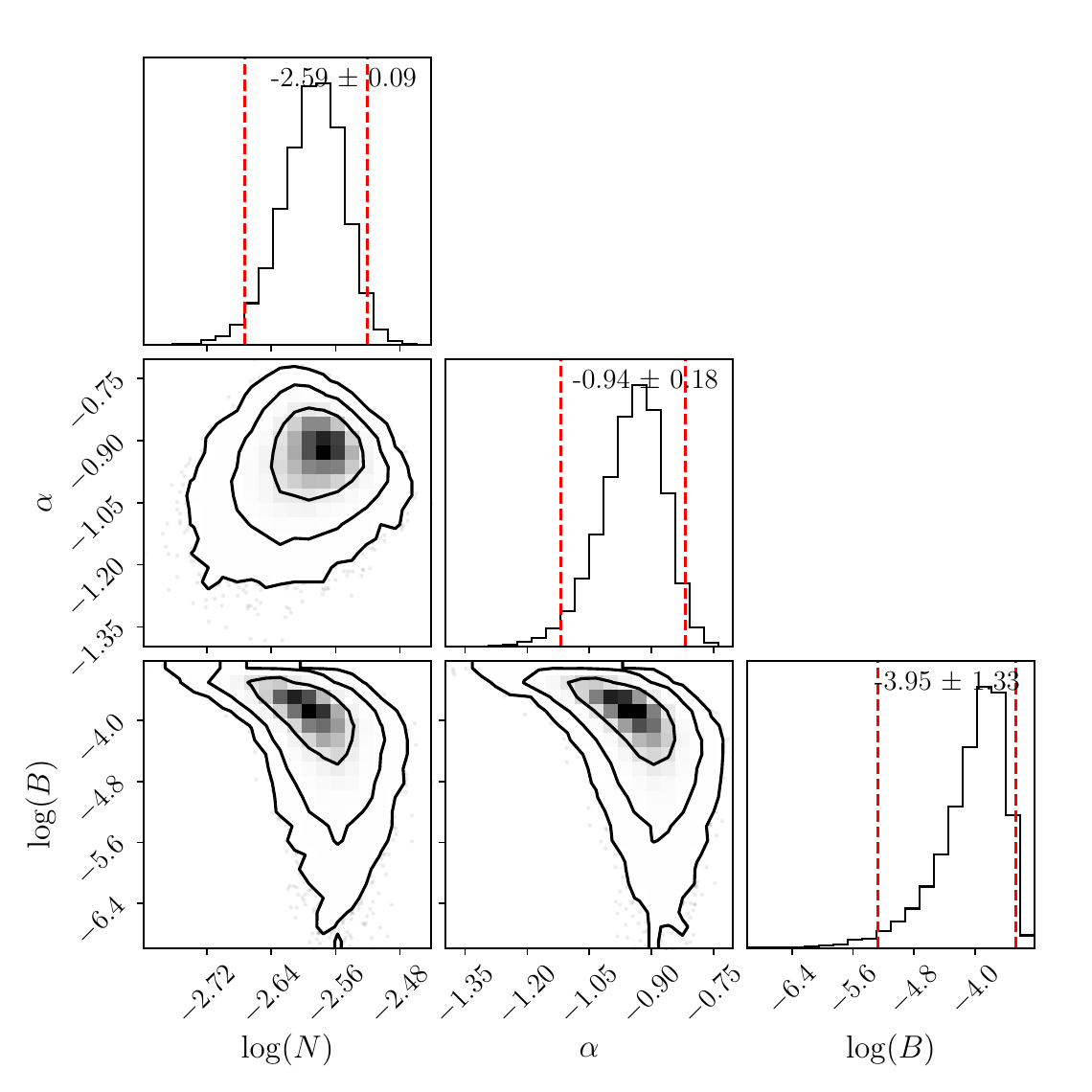}
	\includegraphics[width=0.32\textwidth]{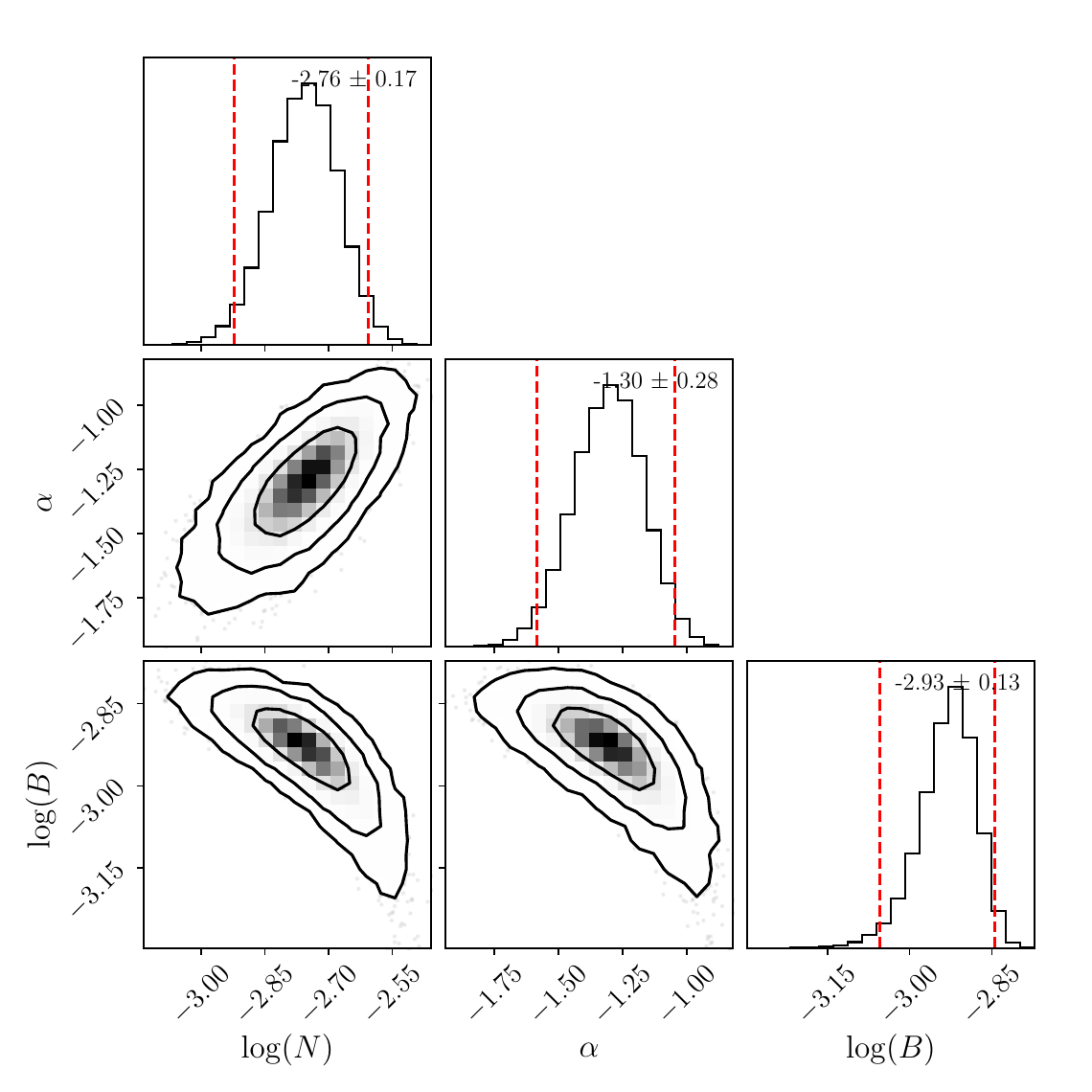}
	\includegraphics[width=0.32\textwidth]{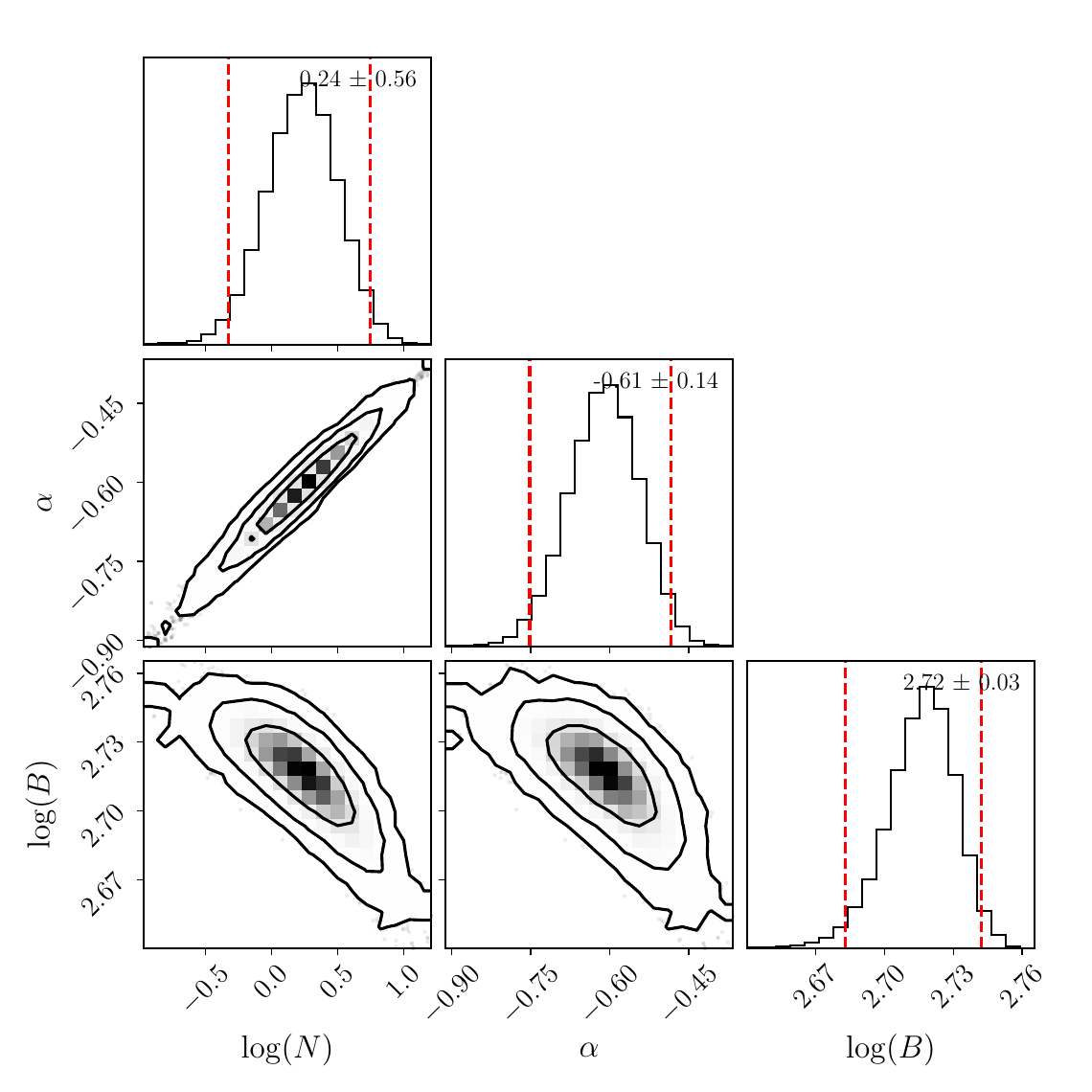}
	\caption{Fitting results of the PDS for GRB~211211A (left), GRB~060614 (middle), and Sw~J1644+57 (right). In the upper panel, the black solid lines represent the PDS generated by LSP, while the red lines show posterior samples from the PSD-fitting MCMC chain. The lower panels display the corner plots illustrating the posterior distributions of the parameters ${\rm{log}} N$, $\alpha$, and ${\rm{log}} B$, with the $3\sigma$ uncertainties for each parameter indicated.}
	
	\label{fig4}
\end{figure*}

Based on the detrended light-curve,
the PDS obtained for GRB~211211A, GRB~060614, and Sw~J1644+57 are showed in the upper panels of Figure~\ref{fig4}.
We fit the PDS using a power-law model, supplemented with a Poisson noise, i.e. (\citealp{2016A&A...589A..98G,2024ApJ...972..190Z}):
\begin{equation}
 P = Nf^{\alpha} + B.
\end{equation}
The PDS fitting is based on a Bayesian approach (\citealp{2016A&A...589A..98G}), 
utilizing a Markov Chain Monte Carlo (MCMC) algorithm, e.g., the Python package \texttt{emcee} (\citealp{2013PASP..125..306F}).
The fitting results for the PDS are showed as red lines in the upper panels of Figure~\ref{fig4}
and reported in Table~\ref{tb1}.
In Figure~\ref{fig4}, the bottom panels is the corner of the parameter distribution from MCMC sampling.
The power-law indices of the PDS are $\alpha = -0.94 \pm 0.18$, $-1.30 \pm 0.28$, and $-0.61 \pm 0.14$ for GRB~211211A, GRB~060614, and Sw~J1644+57, respectively. Based on the derived power-law indices, the exponential factor $\beta$, which characterizes the fluctuations for $\mathrm{d}M/\mathrm{d}E$, is found to be $\sim -1$.

\section{Conclusion and Discussion} \label{sec4}
This study explores potential origins of the fluctuation characteristics observed during the decay phases of the light curves for GRB~211211A, GRB~060614, and Sw J1644+57.
During the tidal disruption process,
a number of tidal disruption debris is ejected and
the bound debris subsequently falls back to form an accretion disk.
The mass distribution ${\rm d}M/{\rm d}E$ of these bound debris moving along Keplerian orbits is influenced by various factors such as self-gravity, shock heating, non-uniform density distributions within the star, and fluid turbulence (\citealp{2015ApJ...808L..11C,2021ApJ...922..168N,2021ApJ...923..184N,2023MNRAS.526.2323F}).
The ${\rm d}M/{\rm d}E$ for the ejected debris is modeled with a fluctuating PDS, following a $\propto f_{\rm E}^{\beta}$ dependence, where $f_{\rm E}\propto 1/E$.
It is shown that the jet power effectively reverses the general fluctuating characteristics of the fallback process
with short timescale variabilities suppressed.
In addition, the case with $\beta \sim -1$ for the fluctuation in the fallback process
well reproduces the observed variabilities in the light-curves,
e.g., GRB~211211A, GRB~060614, and Sw~J1644+57.
The PDS analysis on the observations also reveals $\beta \sim -1$.
We further examin the dependence of the ${\rm d}M/{\rm d}E$ fluctuation model on other parameters and find that the characteristics of the light curve fluctuations depended solely on the index $\beta$.

Additionally, stream–stream and stream–disc interactions during the early formation of the accretion disk are considered another potential source of light curve variability (\citealp{2022MNRAS.510.1627A}), particularly in the earliest stages of the process. Our work demonstrates that after the tidal disruption of a compact or main-sequence star, self-gravity and turbulence within the debris lead to mass fluctuations during the fallback process. These fluctuations imprint on the relativistic jet and manifest as variability in the observed light curves. Our analysis also reveal that the fluctuation properties during the debris fallback process follow a PDS of the form $\propto f^{-1}$.

\emph{Acknowledgments}
{We thank Ziqi Wang and Xiao Li for their helpful discussion and an anonymous reviewer for
	providing constructive feedback. This work is supported by the National Natural Science Foundation of China
(grant Nos. 12273005 and 12133003), the Guangxi Science Foundation (grant Nos. 2018GXNSFFA281010), and China Manned Spaced Project (CMS-CSST-2021-B11).}

\appendix
\section*{Appendix: Data Processing Details}

Here, we provide additional details regarding the data processing. To extract the decay segments of the light curves for the three sources, we first apply a broken power law (BPL) fitting. The BPL function is defined as
\begin{equation}
	f(t)=
	\begin{cases}
		a_1\,t^{b_1}, & \text{if } t < t_\mathrm{break}, \\
		a_1\,t_\mathrm{break}^{(b_1-b_2)}\,t^{b_2}, & \text{if } t \geq t_\mathrm{break},
	\end{cases}
	\tag{A1}
	\label{A1}
\end{equation}
where $t_\mathrm{break}$ denotes the break time in the light curve. For the initial fitting, we set $b_1 = 0$; the resulting parameters are summarized in Table~\ref{tb2}, and the corresponding fitting curves are shown as dashed lines in Figure~\ref{fig:appendix_fig1}.
Next, Equation~\ref{A1} is applied again to specifically target the $-5/3$ power-law decay segments by setting $b_1 = -5/3$. The results of this second fitting step are also summarized in Table~\ref{tb2}, with the fitted curves represented as solid lines in Figure~\ref{fig:appendix_fig1}.
The $-5/3$ power-law decay fitting curve is then adopted to define the overall trend of the light curves. To detrend the data, the extracted $-5/3$ decay segments are divided by the defined trend, yielding the detrended light curves. 
For the source Sw~J1644+57, since the $-5/3$ power-law decay segment is already clearly evident, we directly set $b_1 = 0$ and $b_2 = -5/3$ in Equation~\ref{A1} to extract the decay segment without additional fitting (\citealp{2013ApJ...767..152Z}).

\begin{table}[ht]
	\centering
	\caption{Fitting parameters of the light curves}\label{tb2}
	\begin{tabular}{c|ccc|ccc}
		\hline\hline
		Event & \multicolumn{3}{c|}{Fit 1} & \multicolumn{3}{c}{Fit 2} \\
		\cline{2-7}
		& $t_\mathrm{break}$ & $a_1$ & $b_2$ & $t_\mathrm{break}$ & $a_1$ & $b_2$ \\
		\hline
		GRB~211211A & $21.57 \pm 0.51$\,[s] & $14.77 \pm 0.25$\,[$10^{-7}$] & $-1.92 \pm 0.10$ & $50.38 \pm 1.07$\,[s] & $23.21 \pm 0.16$\,[$10^{-5}$] & $-4.00 \pm 0.35$ \\
		GRB~060614  & $47.70 \pm 0.68$\,[s] & $16.83 \pm 0.21$\,[$10^{-8}$] & $-1.98 \pm 0.10$ & $99.94 \pm 4.20$\,[s] & $98.98 \pm 0.94$\,[$10^{-6}$] & $-4.68 \pm 0.16$ \\
		Sw~J1644+57 & $11.87 \pm 0.16$\,[day] & $23.44 \pm 0.24$\,[$10^{-11}$] & - & - & - & - \\
		\hline
	\end{tabular}
\end{table}

	\clearpage
	
	\clearpage
	\bibliographystyle{aasjournal}
	\bibliography{sample631}

\renewcommand{\thefigure}{A\arabic{figure}}
\setcounter{figure}{0}
\begin{figure}[htbp]
	\centering
	\includegraphics[width=0.49\textwidth]{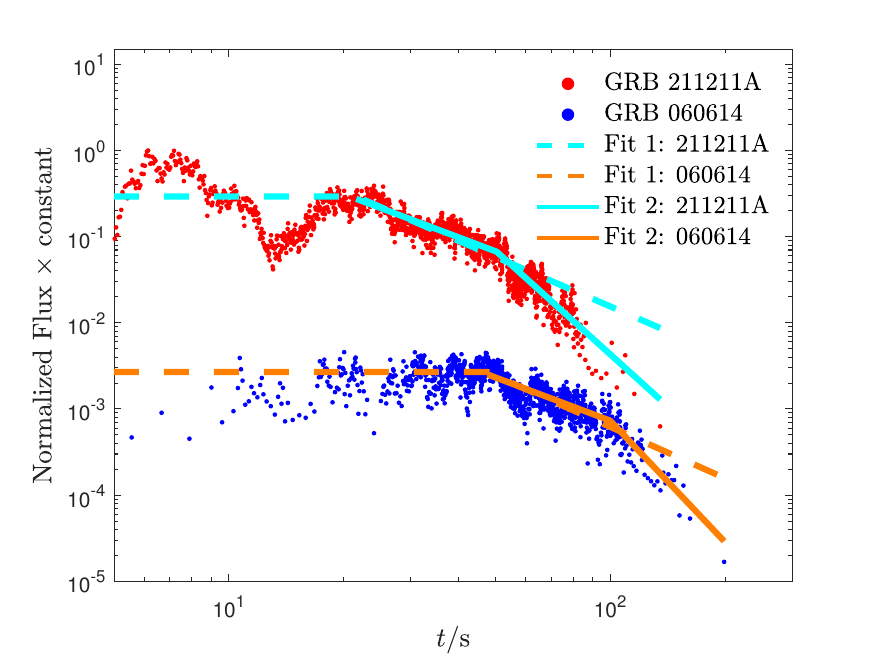}
	\includegraphics[width=0.49\textwidth]{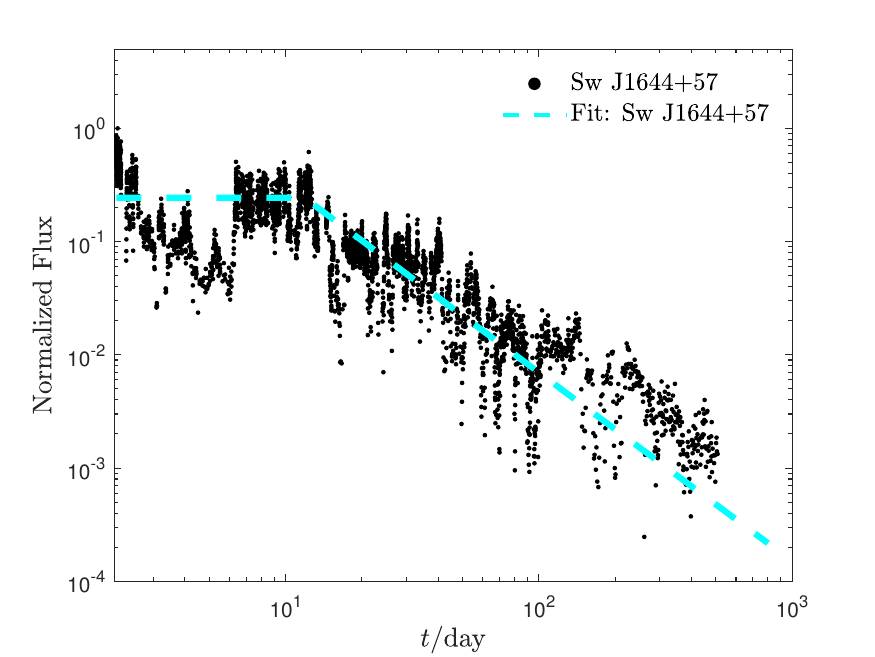}
	\caption{Fitting results for different light curves. In the left panel, the red and blue points represent the normalized light curves of GRB~211211A and GRB~060614, respectively. The cyan dashed and solid lines correspond to the first and second step fits for GRB~211211A, respectively, while the orange dashed and solid lines correspond to the first and second step fits for GRB~060614. To prevent overlap, the normalized curves are scaled by factors of 1 (red), $10^{-2}$ (blue). In the right figure, the black dots represent the light curve of Sw1644+57, and the cyan solid line shows its fitted curve.}
	\label{fig:appendix_fig1}
\end{figure}

\end{document}